\documentclass[10pt]{iopart}
\usepackage{iopams}
\usepackage{graphicx}
\usepackage{dcolumn}
\usepackage{bm}
\usepackage[colorlinks = true, allcolors = blue]{hyperref}
\usepackage{float}

\begin{document}

\title[Algebraic structure underlying different bases of the 9D MICZ-KP]{Algebraic structure underlying spherical, parabolic and prolate spheroidal bases of the nine-dimensional MICZ-Kepler problem}

\author{Dai-Nam Le$^{1,2,3,4}$, Van-Hoang Le$^{5}$ }
 \address{$^1$Department of Theoretical Physics, Faculty of Physics and Engineering Physics, University of Science, 227~Nguyen Van Cu Street,~District 5,~Ho Chi Minh City,~Vietnam}
 \address{$^2$Vietnam National University, Ho Chi Minh City, Vietnam}
 \address{$^3$Atomic~Molecular~and~Optical~Physics~Research~Group, Advanced Institute of Materials Science, Ton Duc Thang University, Ho~Chi~Minh~City, Vietnam}
 \address{$^4$Faculty of Applied Sciences, Ton Duc Thang University, Ho~Chi~Minh~City,~Vietnam}
 \address{$^5$Department of Physics, Ho Chi Minh City University of Education, 280 An Duong Vuong Street, District 5, Ho Chi Minh City, Vietnam}
 \ead{ledainam@tdtu.edu.vn (Dai-Nam Le)} 
 \ead{hoanglv@hcmue.edu.vn (Van-Hoang Le)}

\date{\today}

\begin{abstract}%
The nonrelativistic motion of a charged particle around a dyon in $(9+1)$ spacetime is known as the nine-dimensional MICZ-Kepler problem. This problem has been solved exactly by the variables-separation method in three different coordinate systems, spherical, parabolic, and prolate spheroidal. In the present study, we establish a relationship between the variable separation and the algebraic structure of $SO(10)$ symmetry. Each of the spherical, parabolic, or prolate spheroidal bases is proved to be a set of eigenfunctions of a corresponding nonuplet of algebraically-independent integrals of motion. This finding also helps us establish connections between the bases by the algebraic method. This connection, in turn, allows calculating complicated integrals of confluent Heun, generalized Laguerre, and generalized Jacobi polynomials, which are important in physics and analytics. 
\end{abstract}

\noindent{\it MICZ-Kepler, $SO(8)$ monopole, nine-dimensional space, spherical basis, parabolic basis, prolate spheroidal basis, symmetry, separability, interbasis transformation.\/}

\maketitle

\section{Introduction}
The Kepler problem, describing both the microscopic motion of an electron in a hydrogen atom and the macroscopic motion of planets in the solar system, is one of the most fundamental problems in classical and quantum mechanics. Within the framework of nonrelativistic quantum mechanics, Zwanziger \cite{zwanziger1968exactly}, McIntosh and Cisneros \cite{mcintosh1970degeneracy} had independently investigated the influence of the Dirac magnetic monopole \cite{Dirac1931} on the Kepler problem in the 1960s and, as a result, established for the first time the so-called McIntosh-Cisneros-Zwanziger (MICZ-) Kepler problem.

Analogous to the generalization of the Kepler problem to multidimensional spacetime \cite{Nieto1979,MLADENOV1985,AQUILANTI1997,al1998hydrogen}, the MICZ-Kepler problem has also been generalized from $(3+1)$-spacetime to $(N+1)$-spacetime, where $N=5$ \cite{Iwai1990,Iwai1996,Mardoyan1997}, $N=9$ \cite{le2009hidden,le2011non}, or for the arbitrary spatial dimensionality $N$ \cite{Meng2007,Meng2008,Meng2010,Meng2011}. Among these, the (3, 5, 9)-dimensional MICZ-Kepler problems are more interesting regarding their dual relations with (4, 8, 16)-dimensional isotropic harmonic oscillators \cite{Barut1979,Mardoyan1997,le2009hidden,le2011non} via the Kustaanheimo-Stiefel \cite{KS1965}, Hurwitz \cite{kibler1986hydrogen,davtyan1987generalized,Le1991}, and generalized Hurwitz transformations \cite{Le1993}. Instead of the gauge field of $U(1)$ Dirac monopole \cite{Dirac1931} in the three-dimensional MICZ-Kepler problem, the electron interacts with a dyon via the gauge fields belonging to $SU(2)$ Yang monopole \cite{Yang1978} and $SO(8)$ monopole \cite{grossman1984solutions,pedder2008berry,le2011non} in five- and nine-dimensional ones, respectively. Interestingly, the existence of these monopole fields is the direct consequence of some fundamental mathematics objects like the complex $\mathbb{C}$, quaternion $\mathbb{H}$, and octonion $\mathbb{O}$ normed division algebras \cite{baez2002octonions} or $S_1 \hookrightarrow S_3 \rightarrow S_2$, $S_3 \hookrightarrow S_7 \rightarrow S_4$ and $S_7 \hookrightarrow S_{15} \rightarrow S_{8}$ Hopf maps \cite{hopf1931abbildungen,hopf1935abbildungen}. Moreover, due to the Hopf maps, the Dirac, Yang, and $SO(8)$ monopoles have also been found in condensed matter physics via two-, four-, and eight-dimensional quantum Hall effects \cite{Laughlin1983,zhang2001four,bernevig2003eight}. Notably, in the most recent studies on particle physics in 2018 and 2020 \cite{FUREY2018,Boyle2020}, the last generation of norm division algebra $\mathbb{O}$ has also been used to generate three full generation-structure of the Standard model and build a noncommutative geometry to answer the fundamental question of unifying gravity and particles physics.  These results make the nine-dimensional MICZ-Kepler problem noticeable again. 

In another aspect, the multidimensional MICZ-Kepler problems have been a fascinating research subject for the mathematical physics community for many decades as they are rare quantum mechanical systems with monopole that still exhibit the (maximal) superintegrability. Many works have analytically and algebraically examined the five-dimensional MICZ-Kepler problem in the last two decades \cite{Mardoyan1997,Mardoyan1999,Mardoyan2000,Mardoyan2003,Pletyukhov1999,Marquette2010,Marquette2012,Hoque2017}. Simultaneously, after first being introduced a decade ago, the nine-dimensional MICZ-Kepler problem (9D MICZ-KP) \cite{le2009hidden,le2011non} has also been examined up to now in various aspects such as its (dynamical) symmetry \cite{le2011so,phan2012generalized}, algebraic solutions \cite{le2011so}, analytical solutions with the wavefunctions in spherical \cite{nguyen2015exact}, parabolic and prolate spheroidal coordinates \cite{le2019para}, and also its superintegrability \cite{phan2018super}. Also, in Ref. \cite{le2019para}, the interbasis connection has been constructed between the parabolic and spherical bases. Unfortunately, the parallel connection between the prolate spheroidal and spherical bases cannot be built in an analytical approach since it required a direct integration between confluent Heun \cite{Kereselidze2016}, generalized Laguerre, and generalized Jacobi polynomials \cite{gradshteyn2014table}. This integration has not been analytically calculated yet. Therefore, we expect to solve this remained open problem by another approach that profoundly understands the algebraic structure of each basis of the 9D MICZ-KP.

In this work, the symmetries responsible for spherical, parabolic, and prolate spheroidal bases are investigated under the algebraic approach to understand which integral of motion mainly characterizes the separable coordinate system. In particular, we introduce nonuplets of the algebraically independent integrals of motion for the 9D MICZ-KP corresponding to the separation variables in the spherical, parabolic, and prolate spheroidal coordinates. We then explicitly construct a connection between the prolate spheroidal and spherical bases using the algebraic structure behind these bases. 

\section{Symmetry and superintegrability of the 9D MICZ-KP}\label{sec:2}

In this Section, we first revisit the main results of the work \cite{phan2018super} for the nonuplet of algebraically independent integrals of motion of the 9D MICZ-Kepler problem and then construct the last member of them in three versions. Their application is shown in the next Section.

The time-independent wavefunction $\Psi\left(\bm{r}, \phi \right)$ describing the bound motion ($E < 0$) of a test charge under the presence of the self-dual $SO(8)$ monopole field is governed by the following dimensionless Schr\"odinger equation: \footnote{In this paper, the Latin indices run from 1 to 8 while the Greek ones run from 1 to 9. The Einstein summation convention is used throughout the paper.}
\begin{equation}\label{eqn:Schr}
\left\{ \frac{1}{2} \hat{\bm{\pi}} \cdot \hat{\bm{\pi}} + \frac{\hat{Q}^2}{8 r^2} - \frac{Z}{r} \right\} \Psi\left(\bm{r}, \phi\right) = E \Psi\left(\bm{r}, \phi\right). 
\end{equation}
Here, $Z$ is the electric charge, and $\hat{Q}_{ij}$ are operators describing the $SO(8)$ monopole charges. $\hat{Q}^2 = \hat{Q}_{ij} \hat{Q}_{ij} \; (1 \leq i < j \leq 8)$ is the Casimir operator.  The generalized momentum operators are defined as
\begin{equation}
\label{eqn:pi}
\hat{\bm{\pi}} = \left( - \imath \partial _j + A_k (\bm{r}) \hat{Q}_{kj} , - \imath \partial _9 \right), \quad j,k = 1, 2, \ldots, 8,
\end{equation}
where the monopole vector field is given by
\begin{equation}\label{eqn:A}
A_k (\bm{r}) = \frac{x_k}{r(r+x_9)}.
\end{equation}

Noticeably, $\hat{Q}_{ij}$ are differential operators in terms of variables $\phi _s \, (s=0,1,\ldots,6)$, denoted as $\phi$ for short. Hence, the wavefunction $\Psi\left(\bm{r}, \phi \right)$ depends not only on coordinates of the real space ($\mathbf{r} \in \mathbb{R}^9$) but also on seven additional angles $(\phi _6, \phi _5, \ldots, \phi _0) \in \left[ 0, \pi \right]^6 \times \left[0, 2 \pi \right] $ of a unit sphere $S^7$ arisen from the generalized Hurwitz transformation $\mathbb{R}^{16} \rightarrow \mathbb{R}^9 \times S_7$, connecting a 16D isotropic harmonic oscillator with the 9D Kepler Coulomb problem. We can intuitively understand that each point $\mathbf{r} = (x_1, x_2, \ldots, x_9)$ on the real 9D space contains an abstract space characterized by a seven-dimensional unit sphere $S^7$. 
Since the operator $\hat{Q}^2$ commutes with the Hamiltonian in \Eref{eqn:Schr}, the wavefunction $\Psi$ also satisfies the equation
\begin{equation}\label{eqn:Q}
\hat{Q}^2 \Psi = Q (Q + 6) \Psi.
\end{equation}
Here, the quantum number $Q$ is an integer because angles $\phi _s$ $(s=0,1,\ldots,6)$ have to close the abstract monopole space $S_7$.

The terms $A_k (\bm{r}) \hat{Q}_{kj}\, (j=1,2,\dots,8)$ present the $SO(8)$ monopole interaction, in which monopole operators $\hat{Q}_{kj}$  act on the abstract space only and does not affect the real spatial part of the wavefunction $\Psi( \mathbf{r}, \phi)$. To clearer observe the influences of the $SO(8)$ monopole on the 9D Kepler Coulomb system, we expand the Hamiltonian of the 9D MICZ-KP into a more specific form
\begin{equation}\label{eqn:hamil}
\hat{H} = - \frac{1}{2} \Delta _{\mathbb{R}^9} + \frac{1}{2 r (r + x_9)} \hat{L}_{jk} \hat{Q}_{jk} + \frac{1}{4 r (r+x_9)} \hat{Q}^2 - \frac{Z}{r},
\end{equation}
where $\hat{L}_{jk} = - \imath (x_j \partial _k - x_k \partial _j)$ are projectors of the angular momentum on the $S^7$ sphere $x_1^2 + x_2^2 + \dots + x_8^2 = 1$ of the real space $\mathbb{R}^9$.
The coupling term $\hat{L}_{jk} \hat{Q}_{jk}$ is similar to the term describing the spin-orbital coupling; thus, we refer to the $SO(8)$ monopole interaction as \lq\lq{}isospin\rq\rq{} interaction. This $\hat{L} \cdot \hat{Q}$ coupling term between a test charge and the $SO(8)$ monopole constraints the form of orthogonal coordinate systems to separate the Schr\"odinger equation of the 9D MICZ-KP. Before going on details of the  algebraic structure of the 9D MICZ-KP, we note that both $\hat{L}_{jk}$ and $\hat{Q}_{jk}$ are antisymmetric operators forming a closed $SO(8)$ algebra.

According to Ref. \cite{phan2012generalized}, there are 54 integrals of motion namely generalized angular momentum and Laplace-Runge-Lenz vector operators \footnote{Here, we use the notations: $[\hat A, \hat B ]_{\pm}={\hat A} {\hat B}\pm {\hat B} {\hat A}$.}
\begin{eqnarray}
\hat{\Lambda}_{\mu \nu} && = x _{\mu} \hat{\pi}_{\nu} - x _{\nu} \hat{\pi}_{\mu} + \imath r^2 \left[ \hat{\pi} _{\mu}, \hat{\pi} _{\nu} \right]_{-}, \label{eqn:lambda}  \\
\hat{M} _{\nu} && = \frac{1}{\sqrt{- 2 \hat{H}}} \left\{ \frac{1}{2} \left[ \hat{\pi}_{\mu}, \hat{\Lambda}_{\mu \nu} \right]_{+} + \frac{Z x_{\nu}}{r} \right\} \label{eqn:M}.
\end{eqnarray}
These integrals of motion satisfies the $SO(10)$ algebra and consequently, the 9D MICZ-KP is an $SO(10)$ symmetric system \cite{phan2012generalized,phan2018super}. Additionally, the integrability of the 9D MICZ-KP arises from the existence of nonuplets of algebraically independent integrals constructed from these integrals of motion \cite{phan2018super}. In this study, we denote these nine operators as $\hat{D}_1^2, \hat{D}_2^2, \ldots \hat{D}_9^2$, where superscript $2$ on these operators emphasizes the quadratic form in terms of momentum. Trivially, the first member of the nonuplets is the Hamiltonian $\hat{H} \equiv \hat{D}_1^2$. The next seven members is constructed as follows
\begin{equation}\label{eqn:Dm}
\hat{D}_m^2 = \sum _{1 \leq j < k \leq m}  \hat{\Lambda}_{jk}^2 = \sum _{1 \leq j < k \leq m}  \left(\hat{L}_{jk} + \hat{Q}_{jk} \right)^2, \quad m = 2, 3, \ldots, 8
\end{equation}
which is actually second-order Casimir operators of subgroups $SO(m) \subset SO(10)$. All operators \Eref{eqn:Dm} are quadratic in terms of momentum and commute with the Hamiltonian $[ \hat{D}_m^2, \hat{H}]_{-} = 0$. Also, according to the fact that $SO(2) \subset SO(3) \subset \cdots \subset SO(8)$, these operators $\hat{D}_m^2$ are all independent and also commute with each other $[\hat{D}_m^2, \hat{D}_n^2 ]_{-} = 0$. The conversation of $\hat{D}_2^2, \hat{D}_3^2, \ldots, \hat{D}_8^2$ is the direct consequence of the $\hat{L} \cdot \hat{Q}$ coupling interaction between the orbital angular momentum of a test charge and the \lq\lq{}isospin\rq\rq{} generators of the $SO(8)$ monopole. In consistency with the previous works, we use also new notations: $\hat J^2 \equiv \hat D_8^2$, $\hat j_5^2 \equiv \hat D_7^2$, $\hat j_4^2 \equiv \hat D_6^2$, $\hat j_3^2 \equiv \hat D_5^2$, $\hat J_2^2 \equiv \hat D_4^2$, $\hat j_1^2 \equiv \hat D_3^2$, and $\hat j_0^2 \equiv \hat D_2^2$ for the operators \Eref{eqn:Dm}.

Now, we look for the last member $\hat{D}_9^2$ of the nonuplets of first integrals. This integral of motion must obey two following properties:
\begin{itemize}
    \item commute with all of the above operators $\hat{D}_1^2, \hat{D}_2^2, \ldots, \hat{D}_8^2$;
    \item be a second-order operator in term of momentum. 
\end{itemize}
After carefully investigation, we found that there are only three choices for the last member $\hat{D}_9^2$ which are total generalized angular momentum $\hat{\Lambda} ^2 = \sum _{1 \leq \mu < \nu \leq 9} \Lambda_{\mu \nu}^2$, the component $\hat{M}_9$ of the Laplace-Runge-Lenz vector and their linear combination with a real parameter $\beta$:
\begin{equation}\label{eqn:linear}
\hat{\Lambda}^2_M \equiv \Lambda ^2 + \beta \hat{M}_9 .
\end{equation}
These operators $\hat{\Lambda} ^2$, $\hat{M}_9$ and $\hat{\Lambda}^2_M$ commute with all of $\hat{D}_1^2, \hat{D}_2^2, \ldots, \hat{D}_8^2$ as the results of $SO(10)$ symmetry of the 9D MICZ-KP while their explicit forms in Cartesian coordinates  
\begin{eqnarray}\label{eqn:Lambda}
\Lambda ^2 &&  = - r^2 \Delta _{\mathbb{R}^9} + 7 \left( \bm{r} \cdot \bm{\nabla} \right) + \left(\bm{r} \cdot \bm{\nabla} \right)^2 + \frac{r \hat{L}_{jk} \hat{Q}_{jk}}{(r + x_9)}  + \frac{r \hat{Q}^2}{2 (r+x_9)}  ,
\end{eqnarray}
and
\begin{eqnarray}\label{eqn:M9}
\hat{M}_9 = \frac{1}{\sqrt{-2 \hat{H}}} && \left\{ x_9 \Delta _{\mathbb{R}^9} - \left( \bm{r} \cdot \bm{\nabla} + 4 \right) \partial _9 + \frac{Z x_9}{r} \right.\nonumber\\
&& \quad \quad \left. + \frac{r - x_9}{2 r (r+x_9)}\hat{L}_{jk} \hat{Q}_{jk} + \frac{r-x_9}{4 r (r+x_9)} \hat{Q}^2  \right\},
\end{eqnarray}
are obviously of second-order operators in term of momentum. 

The next Section will show that three different cases of choosing $\hat{D}_9^2$ correspond to three different separating coordinate systems, namely spherical ($\hat{D}_9^2 =\hat{\Lambda} ^2$), parabolic ($\hat{D}_9^2 =\hat{M}_9$), and prolate spheroidal ($\hat{D}_9^2 = \hat{\Lambda}_M^2$).

\section{Separable coordinate systems and algebraic structure}\label{sec:3}

In this Section, we provide the variable separation of the 9D MICZ-Kepler problem in three coordinate systems (spherical, parabolic, and prolate spheroidal) based on the work \cite{nguyen2015exact,le2019para} and then reveal the algebraic structure behind these separations.  

The Hamiltonian \Eref{eqn:hamil} contains the $SO(8)$ monopole interaction via the coupling term $\hat{L}_{jk} \hat{Q}_{jk}$. This form of the Hamiltonian suggests that we separate the Schr\"odinger equation of the 9D MICZ-KP using a curvilinear orthogonal coordinate system, which parametrizes the $S^7$ sphere $x_1^2 + x_2^2 + \ldots + x_8^2 = 1$ by seven angles $(\varphi _0, \varphi _1, \ldots, \varphi_6)$, denoted $\varphi$ for short. Such a coordinate system can be written as   
\begin{equation}\label{eqn:coord}
\begin{array}{lll}
x_9 &=& z,\\
x_8 &=& \rho \cos \varphi _6,  \\
x_7 &=& \rho \sin \varphi _6 \cos \varphi _5 ,  \\
x_6 &=& \rho \sin \varphi _6 \sin \varphi _5 \cos \varphi _4 , \\
x_5 &=& \rho \sin \varphi _6 \sin \varphi _5 \sin \varphi _4 \cos \varphi _3 , \\
x_4 &=& \rho \sin \varphi _6 \sin \varphi _5 \sin \varphi _4 \sin \varphi _3 \cos \varphi _2 , \\
x_3 &=& \rho \sin \varphi _6 \sin \varphi _5 \sin \varphi _4 \sin \varphi _3 \sin \varphi _2 \cos \varphi _1 , \\
x_2 &=& \rho \sin \varphi _6 \sin \varphi _5 \sin \varphi _4 \sin \varphi _3 \sin \varphi _2 \sin \varphi _1 \cos \varphi _0 , \\
x_1 &=& \rho \sin \varphi _6 \sin \varphi _5 \sin \varphi _4 \sin \varphi _3 \sin \varphi _2 \sin \varphi _1 \sin \varphi _0 ,
\end{array}
\end{equation}
whereas  $\left( \varphi _6, \varphi _5, \ldots, \varphi _0 \right) \in \left[ 0, \pi \right]^6 \times \left[0, 2 \pi \right]$ are canonical parameters of the $S^7$ sphere $x_1^2 + x_2^2 + \ldots + x_8^2 = \rho^2$; $\left(z, \rho \right) \in \left[-\infty, +\infty \right] \times \left[0, +\infty \right]$ are two real variables which can be transformed further for different coordinate systems.

Within the coordinate transformation \Eref{eqn:coord}, the Hamiltonian \Eref{eqn:hamil} can be read as
\begin{equation}\label{eqn:hamil-coord}
\hat{H} = - \frac{1}{2} \Delta_{z \rho} + \frac{\hat{L}^2(\varphi)}{4r(r - z)} + \frac{\hat{J}^2(\varphi,\phi)}{4r (r+z)} - \frac{Z}{r},
\end{equation}
where $r=\sqrt{z^2+\rho^2}$; $\Delta _{z \rho}$ is Laplace-Beltrami operator of $(z, \rho)$ subspace; $\hat{L}^2 = \sum _{1 \leq j < k \leq 8} \hat{L}_{jk} ^2$ depends on angles $\left( \varphi _6, \varphi _5, \ldots, \varphi _0 \right)$ only. 
Regarding the independence of  $\hat{L}^2$ and $\hat{J}^2$ on $(z, \rho)$, the 9D MICZ-KP wavefunction can be separated as $\Psi (z, \rho, \varphi, \phi) = \psi (z, \rho) \mathcal{D} \left( \varphi, \phi \right)$ in which $D \left( \varphi, \phi \right)$ is an eigenfunction of $\hat{L}^2$ and $\hat{J}^2$, i.e., satisfies the below equations \cite{nguyen2015exact}
\begin{equation}\label{eqn:Dfunction}
\left\{
\begin{array}{l}
\hat{L}^2 \mathcal{D} \left( \varphi, \phi\right) = L (L+6) \mathcal{D} \left( \varphi, \phi \right) \\
\hat{J}^2 \mathcal{D} \left( \varphi, \phi \right) = J (J+6) \mathcal{D} \left( \varphi, \phi \right).
\end{array}
\right. 
\end{equation}
The explicit expression of $\mathcal{D}_{L,Q,J,j_5,\ldots,j_1,m_j} (\varphi, \phi)$ with the eigenvalues $L$, $J$, $j_5$, $j_4$, $j_3$, $j_2$, $j_1$, $m_j$ can be found in Ref. \cite{nguyen2015exact}, in which this angular wavefunction is also required to satisfy the following equations
\begin{equation}\label{eqn:DfunctionB}
\left\{
\begin{array}{l}
\hat{j}^2_5 \mathcal{D} \left( \varphi, \phi\right) = j_5 (j_5+5) \mathcal{D} \left( \varphi, \phi \right) \\
\hat{j}^2_4 \mathcal{D} \left( \varphi, \phi \right) = j_4 (j_4+4) \mathcal{D} \left( \varphi, \phi\right) \\
\hat{j}^2_3 \mathcal{D} \left( \varphi, \phi \right) = j_3 (j_3+3) \mathcal{D} \left( \varphi, \phi\right) \\
\hat{j}^2_2 \mathcal{D} \left( \varphi, \phi \right) = j_2 (j_2+2) \mathcal{D} \left( \varphi, \phi\right) \\
\hat{j}^2_1 \mathcal{D} \left( \varphi, \phi \right) = j_1 (j_1+1) \mathcal{D} \left( \varphi, \phi\right) \\
\hat{j}^2_0 \mathcal{D} \left( \varphi, \phi \right) = m_j^2 \mathcal{D} \left( \varphi, \phi\right).
\end{array}
\right. 
\end{equation}

Substituting the separation form of the wave function $\Psi (z, \rho, \varphi, \phi)$ into \Eref{eqn:hamil-coord} and noticing \Eref{eqn:Dfunction}, one may easily show that $\psi (z, \rho)$ obeys the following separated Schr{\"o}dinger equation
\begin{equation}\label{eqn:psi}
\left\{ - \frac{1}{2} \Delta_{z\rho} + \frac{L(L+6)}{4r(r - z)} + \frac{J(J+6)}{4r (r+z)} - \frac{Z}{r} \right\} \psi (z, \rho) = E \psi (z, \rho).
\end{equation}
Henceforth, the variable separability of the equation depends on the transformation of two coordinates $(z, \rho)$.

Back to \Eref{eqn:Dfunction} and \Eref{eqn:DfunctionB}, since the angular wavefunction $\mathcal{D}( \varphi, \phi)$ is an eigenfunction of $\hat{D}_2^2 = \hat{j}^2_0, \hat{D}_3^2 = \hat{j}^2_1, \ldots, \hat{D}_8^2 = \hat{J}^2$, these operators are obviously the first integrals of motion related to the basis set $\psi (z, \rho) \mathcal{D} \left( \varphi, \phi \right)$. Also, \Eref{eqn:psi} is an eigensystem problem related to the Hamiltonian $\hat{D}_1 = \hat{H}$ itself; thus, the explicit form of the last member $\hat{D}_{9}^2$ depends on how to separate \Eref{eqn:psi}. In principle, separating this equation arises another separation constant, which is possibly an eigenvalue of the last operator  $\hat{D}_{9}^2$. This situation suggests a method to define the explicit form of the last integral of motion. We will apply this prediction on spherical, parabolic, and prolate spheroidal bases of the 9D MICZ-KP. 

\subsection{Spherical coordinates}\label{sec:3a}
From Ref. \cite{nguyen2015exact}, \Eref{eqn:psi} has the form in spherical coordinates as
\begin{eqnarray}
&&\left\{ - \frac{1}{2} r^{-8} \partial _r \left( r^8 \partial _r \right) - \frac{1}{2r^2} \left[ \sin ^{-7} \theta \partial _{\theta} \left( \sin ^7 \theta \partial _{\theta} \right) + \frac{L(L+6)}{4 \sin^2 \theta/2} \right. \right. \nonumber\\
&&\quad\quad\quad \quad\quad\quad\quad\quad
\left. \left. + \frac{J(J+6)}{4 \cos ^2 \theta / 2} \right] - \frac{Z}{r} \right\} \psi (r,\theta) = E \,\psi (r, \theta),
\end{eqnarray}
whereas $r = \sqrt{\rho^2 + z^2}$ and $\theta = \arctan \rho / z$.
By the separated wavefunction $\psi (r, \theta) = R(r) \Theta (\theta)$, this equation is separated into two equations
\begin{eqnarray}
\fl \left\{ - \frac{1}{2} r^{-8} \partial _r \left( r^8 \partial _r \right) - \frac{\lambda (\lambda + 7)}{2r^2} - \frac{Z}{r} \right\} R(r) = E R(r) , \label{eqn:R}\\
\fl \left\{ \sin ^{-7} \theta \partial _{\theta} \left( \sin ^7 \theta \partial _{\theta} \right) + \frac{L(L+6)}{4 \sin^2 \theta/2} + \frac{J(J+6)}{4 \cos ^2 \theta / 2} \right\} \Theta (\theta) = \lambda (\lambda + 7) \Theta (\theta) , \label{eqn:theta}
\end{eqnarray}
with the separation constant $\lambda = (L+J)/2, (L+J)/2 + 1, \ldots, n + Q/2$. Solutions of \Eref{eqn:R} and \Eref{eqn:theta} are given in Ref. \cite{nguyen2015exact}, where the wavefunction with quantum numbers is read as
\begin{eqnarray}\label{eqn:psi-sphere}
\Psi ^{spherical}_{n,\lambda,L,Q,J,j_5,\ldots,j_1,m_j} (r, \theta, \varphi, \phi) = && R_{n, \lambda} (r) \Theta _{\lambda, L,Q,J} (\theta) \nonumber\\
&& \times \mathcal{D}_{L,Q,J,j_5,\ldots,j_1,m_j} (\varphi, \phi).
\end{eqnarray}
This is the spherical basis set of the wavefunctions.

On the other hand, we can rewrite the operator $\hat{\Lambda} ^2$ from Cartesian coordinates as in \Eref{eqn:Lambda} into spherical coordinates $(r, \theta)$ as
\begin{equation}\label{eqn:Lambda2}
\hat{\Lambda} ^2 = \sin ^{-7} \theta \partial _{\theta} \left( \sin ^7 \theta \partial _{\theta} \right) + \frac{\hat{L}^2}{4 \sin^2 \theta/2} + \frac{\hat{J}^2}{4 \cos ^2 \theta / 2} ,
\end{equation}
and compare it with the operator on the left-hand side of \Eref{eqn:theta}.
Then, taking the action of operator $\hat{\Lambda} ^2$ on the wave function \Eref{eqn:psi-sphere} and considering \Eref{eqn:Dfunction} and \Eref{eqn:theta}, we can prove that the basis set \Eref{eqn:psi-sphere} satisfies the equation
\begin{equation}\label{eqn:sphere-define}
\hat{\Lambda} ^2 \Psi ^{spherical} _{n,\lambda,L,Q,J,j_5, \ldots, j_1, m_j} = \lambda \left( \lambda + 7 \right) \Psi^{spherical}_{n,\lambda,L,Q,J,j_5, \ldots, j_1, m_j}.
\end{equation}

This equation indicates that the generalized angular momentum described by the operator $\hat{\Lambda}^2$ is conserved under the spherical basis set of the wavefunctions. This circumstance is consistent with the fact that the operator $\hat{\Lambda}^2$ can be chosen as the last member $\hat{D}_9^2$ of our nonuplet of integrals, as shown in Section \ref{sec:2}. Therefore, we can say that the operator $\hat{\Lambda}^2$ is related to the separation of variables in spherical coordinates. Moreover, \Eref{eqn:sphere-define} with the eigenvalue $\lambda (\lambda + 7)$, where $\lambda$ is a non-negative integer, is consistent with the fact that $\hat{\Lambda} ^2$ is the second-order Casimir operator of $SO(9)$ algebra.

\subsection{Parabolic coordinates}\label{sec:3b}
In parabolic coordinates $(u,v)$, whereas $u = r+z$ and $v = r - z$, \Eref{eqn:psi} can be rewritten as
\begin{eqnarray}
&& \left\{ - \frac{2}{u+v} \left( u \partial _u^2 + 4 \partial _u + v \partial _v^2 + 4 \partial _v \right)  + \frac{L(L+6)}{2 v (u+v)} \right. \nonumber\\
&& \quad\quad\quad \quad \left. + \frac{J(J+6)}{2 u (u+v)} - \frac{2 Z}{u+v} \right\} \psi (u, v) = E \psi (u, v).
\end{eqnarray}
As shown in Refs. \cite{phan2018super, le2019para}, this equation is separable on coordinates ($u, v$) with the wavefunction chosen in the form $\psi (u, v)= U(u)V(v)$, where $U(u)$ and $V(v)$ satisfy the following equations:
\begin{eqnarray}
&& \left\{ u \partial _u^2 + 4 \partial _u - J(J+6) u^{-1} + Z/2 + E u / 2 - P \right\} U(u) = 0 \label{eqn:U},\\
&& \left\{ v \partial _v^2 + 4 \partial _v - L(L+6) v^{-1} + Z/2 + E v / 2 + P \right\} V(v) = 0 \label{eqn:V},
\end{eqnarray}
with a separation constant $P$.  Also, \Eref{eqn:U} and \Eref{eqn:V} have been solved analytically in Ref. \cite{le2019para} 
with the separation constant $P$ quantized as
\begin{equation*}
\frac{2 P}{\sqrt{-2E}} = n + \frac{Q}{2} - J - 2 n_p,
\end{equation*} 
with $n_p = 0,1,2, \ldots, n + Q/2 - (L+J)/2$. Then, the basis set of wave functions can be written in the parabolic coordinates as
\begin{eqnarray}\label{eqn:psi-para}
\Psi ^{parabolic} _{n,n_p,L,Q,J,j_5, \ldots, j_1, m_j} (u, v, \varphi, \phi) = && U_{n,n_p,J} (u) V_{n,n_p,L} (v) \nonumber \\
&& \times \mathcal{D}_{L,Q,J,j_5,\ldots,j_1,m_j} (\varphi, \phi). 
\end{eqnarray}

On the other hand, we need to extract the integral of motion corresponding to the separation constant $P$. For this purpose, we first replace $U(u)$ and $V(v)$ in \Eref{eqn:U} and \Eref{eqn:V} by $U(u) V(v)$ and then recombine the two equations with excluding the energy $E$ from the equations. As a result, we obtain an equation 
\begin{eqnarray}\label{eqn:para-P-2}
 \left\{ \frac{2 u v}{u + v}  \right. && \left( \partial _u^2 - \partial _v^2 + 4 v^{-1} \partial _v - 4 u^{-1} \partial _u \right) + \frac{v J(J+6)}{2 u (u+v)} \nonumber\\ 
&&\left. -\frac{u L(L+6)}{2v (u+v)} + \frac{Z (u-v)}{u+v} \right\} U(u) V(v)  = -2P U(u) V(v).
\end{eqnarray}
Interestingly, we note that the operator on the left-hand side of \Eref{eqn:para-P-2} is similar to the Laplace-Runge-Lenz component $\hat{M}_9$ written in parabolic coordinates $(u,v)$ as
\begin{eqnarray}\label{eqn:para-M9}
\hat{M}_9 = \frac{1}{\sqrt{-2 \hat{H}}} && \left\{ \frac{2 u v}{u + v} \left( \partial _u^2 - \partial _v^2 + 4 v^{-1} \partial _v - 4 u^{-1} \partial _u \right) \right. \nonumber\\
&& \left. + \frac{v \hat{J}^2}{2 u (u+v)} - \frac{u \hat{L}^2}{2v (u+v)} + \frac{Z (u-v)}{u+v} \right\}.
\end{eqnarray}
With this observation, we can easily obtain the equation
\begin{equation}\label{eqn:para}
\hat{M}_9 \Psi ^{parabolic} _{n,n_p,L,Q,J,j_5, \ldots, j_1, m_j} = -\frac{2P}{\sqrt{-2E}} \Psi^{parabolic}_{n,n_p,L,Q,J,j_5, \ldots, j_1, m_j},
\end{equation}
which means separating the variables in parabolic coordinates corresponding to the operator $\hat{M}_9$ chosen for the last integral $\hat D^2_9$.

\subsection{Prolate spheroidal coordinates}\label{sec:3c}
We are now applying the similar procedure to prolate spheroidal coordinates $(\xi, \eta)$, defined as \cite{le2019para}
\begin{equation}\label{eqn:xi-eta}
\left\{
\begin{array}{l}
\xi = \frac{ \sqrt{\rho ^2 + z^2} + \sqrt{\rho^2 + (z- a) ^2} }{a}, \\
\eta = \frac{ \sqrt{\rho ^2 + z^2} - \sqrt{\rho^2 + (z - a) ^2} }{a},
\end{array}
\right.
\end{equation}
whereas positive parameter $a$ is the distance between two centers of prolate spheroidal coordinates. It is interesting to notice that this coordinate system becomes the spherical one when $a \to 0$ and the parabolic one when $a \to +\infty$. Changing $\rho,z$ variables by the prolate spheroidal coordinates ($\xi, \eta$), we represent \Eref{eqn:psi} in the form \cite{le2019para}
\begin{eqnarray}\label{eqn:P}
\fl  \left\{ - \frac{1}{ a^2 (\xi ^2 - \eta ^2)} 
\left[ 2(\xi ^2 - 1) \partial _{\xi}^2 + 16 \xi \partial _{\xi} +2(1 -\eta ^2) \partial _{\eta}^2 - 16 \eta \partial _{\eta} \right] - \frac{2Z}{a(\xi + \eta)} \right. \nonumber\\
\fl \quad\quad\left. +\frac{L(L+6) }{a^2(\xi+\eta)(\xi - 1)(1-\eta)} - \frac{J(J+6) }{a^2 (\xi+\eta)(\xi +1)(1+\eta)} 
\right\} \psi (\xi, \eta) = E \psi (\xi, \eta).
\end{eqnarray}

\Eref{eqn:P} is separable by variables $\xi$ and $\eta$ as presented for the first time in Ref. \cite{le2019para}. By setting $\psi (\xi, \eta) = \Xi(\xi) H (\eta)$, this equation is separated into two independent differential equations for $\Xi (\xi)$ and $H (\eta)$ as
\begin{eqnarray}\label{eqn:a1}
 &&\left\{(\xi ^2 - 1) \partial _{\xi}^2 + 8 \xi \partial _{\xi} -\frac{L(L+6)}{2 (\xi - 1)}  
+\frac{ J (J+6) }{2 (\xi + 1)}  \right.\nonumber\\
&&\quad\quad\quad\quad\quad\quad\quad\quad\quad 
\left.+ \frac{1}{2} E a^2 (\xi ^2-1) + Za \xi + K \right\}  \Xi (\xi) =0, \\
 &&\left\{(1 - \eta ^2) \partial _{\eta}^2 - 8 \eta \partial _{\eta}  
- \frac{L(L+6)}{2 (1 - \eta)}  - \frac{ J (J+6) }{2 (1 + \eta )}     \right. \nonumber\\
 && \left. \quad\quad\quad\quad\quad\quad\quad\quad\quad 
 + \frac{1}{2}E a^2 (1- \eta ^2 ) - Z a \eta - K \right\} H (\eta) =0, \label{eqn:b1}
\end{eqnarray}
with a separation constant $K$. These equations have analytical solutions given in Ref. \cite{le2019para} via confluent Heun polynomials, where the separation constant $K$ was found as a function of $n, L, J, Q$ and new quantum number $n_k$, a non-negative integer. The basis set of wavefunctions in prolate spheroidal coordinates is then defined as
\begin{eqnarray}\label{eqn:psi-spheroid}
\Psi ^{spheroidal} _{n,n_{k},L,Q,J,j_5, \ldots, j_1, m_j} (\xi, \eta, \varphi, \phi) = && \Xi_{n,n_{k},L,J} (u) H_{n,n_{k},L,J} (\eta) \nonumber\\
&& \times \mathcal{D}_{L,Q,J,j_5,\ldots,j_1,m_j} (\varphi, \phi).
\end{eqnarray}

On the other hand, excluding $E$ from \Eref{eqn:a1} and \Eref{eqn:b1}, we can combine these two equations into one equation as
\begin{eqnarray}\label{eqn:spheroi-D9}
&& \left\{ \frac{(\xi^2-1)(1-\eta^2)}{\xi^2 - \eta^2} \right.  \left( \partial _{\xi}^2 + \frac{\xi}{\xi^2 - 1} \partial _{\xi} - \partial _{\eta}^2 - \frac{\eta}{1- \eta^2} \partial _{\eta} \right) \nonumber\\
&&\quad \quad\quad+ \frac{L(L+6)}{2} \left( \frac{1}{1-\eta} - \frac{1}{\xi - 1} + \frac{1}{\xi + \eta} \right) \nonumber\\
&&\quad \quad\quad\quad+  \frac{J(J+6)}{2} \left( \frac{1}{1+\eta} + \frac{1}{\xi + 1} - \frac{1}{\xi + \eta} \right) \nonumber\\
&&\quad\quad\quad\quad\quad \left. + \frac{Z a(1+ \xi \eta)}{\xi + \eta} \right\} \Xi (\xi) H (\eta) = - K \; \Xi (\xi) H (\eta).
\end{eqnarray}
This is the eigen-equation for the separation constant $K$. Amazingly, the operator in the left-hand side of this equation is similar to the operator $\hat \Lambda_M^2=\hat \Lambda^2+a\sqrt{-2\hat H}\,\hat M_9$ written in prolate spheroidal coordinates as
\begin{eqnarray}\label{eqn:speroid-D9-2}
\hat{\Lambda }^2_M  = && \frac{(\xi^2-1)(1-\eta^2)}{\xi^2 - \eta^2} \left( \partial _{\xi}^2 + \frac{\xi}{\xi^2 - 1} \partial _{\xi} - \partial _{\eta}^2 - \frac{\eta}{1- \eta^2} \partial _{\eta} \right) \nonumber\\
&& \quad+ \frac{\hat{L}^2}{2} \left( \frac{1}{1-\eta} - \frac{1}{\xi - 1} + \frac{1}{\xi + \eta} \right) + \nonumber\\
&& \quad\quad+  \frac{\hat{J}^2}{2} \left( \frac{1}{1+\eta} + \frac{1}{\xi + 1} - \frac{1}{\xi + \eta} \right) + \frac{Z a(1+ \xi \eta)}{\xi + \eta} .
\end{eqnarray}
Therefore, acting operator $\hat \Lambda^2_M$ on the wavefunction \Eref{eqn:psi-spheroid} and considering \Eref{eqn:spheroi-D9} and \Eref{eqn:Dfunction}, we have the equation:
\begin{equation}\label{eqn:spheroid}
\hat{\Lambda }^2_M \Psi ^{spheroidal}_{n,n_{k},L,Q,J,j_5,\ldots,j_1,m_j} = - K \Psi ^{spheroidal}_{n,n_{k},L,Q,J,j_5,\ldots,j_1,m_j}.   
\end{equation}
This means that the wavefunctions in prolate spheroidal coordinates are eigensolutions of the operator $\hat \Lambda^2_M$. In other words, we can choose the operator  $\hat \Lambda_M^2=\hat \Lambda^2+a\sqrt{-2\hat H}\,\hat M_9$ for the last integrals $\hat D^2_9$ and this choice corresponds to the variables separation in prolate spheroidal coordinates.

\section{Algebraic relations between bases}\label{sec:4}
After constructing the spherical, parabolic, and prolate spheroidal bases, the next important task is to construct transformations between these bases. However, this construction is not trivial. Although the spherical and parabolic bases for the 3D MICZ-KP were first studied by McIntosh and Cisneros in 1970 \cite{mcintosh1970degeneracy}, the connection between these bases has derived analytically and expressed in $SU(2)$ Clebsch-Gordan coefficients in 2005 \cite{Mardoyan2005} only. A similar connection was also built for the 5D MICZ-KP by Mardoyan {\it et al.} \cite{Mardoyan2000} and for 9D MICZ-KP in our recent work \cite{le2019para}. Unfortunately, the connection between the prolate spheroidal and spherical bases is not easy to establish. For the 5D MICZ-KP, Madoyan {\it et al.} \cite{Mardoyan2000} have expanded the prolate spheroidal basis via the spherical basis; however, the expansion coefficients can not be obtained explicitly but satisfy the recurrent equations. For the 9D MICZ-KP, we can not establish the connection between the bases in prolate spheroidal and spherical coordinates because of the complicated integral between the confluent Heun, generalized Laguerre, and generalized Jacobi polynomials. Fortunately, we can perform this task using the algebraic connections given in the previous section and will show in the following.

\subsection{Matrix elements of $\hat{\Lambda} ^2$ and $\hat{M}_9$}\label{sec:4a}
For further calculations, we derive here some analytical formulae for the matrix elements of the operators $\hat{\Lambda} ^2$ and $\hat{M}_9$ with respect to the spherical basis set \Eref{eqn:psi-sphere}. 

As shown in \Eref{eqn:Lambda2}, the operator $\hat{\Lambda}^2$ in spherical coordinates does not depend on radial variable $r$, on the one hand. On the other hand, this operator depends on angles $\varphi$ and $\phi$ via the operators $\hat L^2 (\varphi)$ and $\hat J^2 (\varphi, \phi)$ only. In turn, these operators are diagonal corresponding to the spherical basis set \Eref{eqn:psi-sphere}. Therefore, we will focus only on the part of the wavefunction dependent on the variable $\theta$ and hence calculate the matrix elements: 
\begin{equation}\label{eqn:matrix-lambdaA}
(\hat{\Lambda} ^2)^{n,L,Q,J}_{ \lambda^{\prime}, \lambda} 
= \langle \Psi ^{spherical}_{n,\lambda^{\prime},L,Q,J,j_5,\ldots,j_1,m_j}| 
         \hat{\Lambda}^2 | \Psi ^{spherical}_{n,\lambda,L,Q,J,j_5,\ldots,j_1,m_j}\rangle,
\end{equation}
with the difference only on $\lambda$ and $\lambda^{\prime}$. From the orthogonality of the spherical basis set and \Eref{eqn:sphere-define}, we have
\begin{equation}\label{eqn:matrix-lambda}
(\hat{\Lambda} ^2)^{n,L,Q,J}_{ \lambda^{\prime}, \lambda} = \lambda (\lambda + 7) \delta _{\lambda ^{\prime} , \lambda} .
\end{equation}

Oppositely, in the case of Laplace-Runge-Lenz component $\hat{M}_9$, the explicit expression \Eref{eqn:para-M9} of this operator in parabolic coordinates depends on angles $\varphi$ and $\phi$ via operators $\hat L^2 (\varphi)$ and $\hat J^2 (\varphi, \phi)$ only while  depends on variables $u$, $v$ via $\partial _u$, $\partial _v$, and $u,v$. In the parabolic basis set, both $\hat L^2 (\varphi)$ and $\hat J^2 (\varphi, \phi)$ are also diagonal; thus, only the part of the parabolic wavefunction dependent on variables $(u,v)$ is considered to calculate the matrix element
\begin{eqnarray}\label{eqn:matrix-M9-para}
(\hat{M}_9)^{n,L,Q,J}_{ n_p^{\prime}, n_p} && =  \langle \Psi ^{parabolic}_{n,n_p^{\prime},L,Q,J,j_5,\ldots,j_1,m_j}| 
         \hat{M}_9 | \Psi ^{parabolic}_{n,n_p,L,Q,J,j_5,\ldots,j_1,m_j}\rangle \nonumber\\
         && = - \left( n + \frac{Q}{2} - J - 2 n_p \right) \delta _{n_p ^{\prime} n_p} ,
\end{eqnarray} 
where the orthogonality of the parabolic basis set as well as \Eref{eqn:para} are used for simplification.
 
To calculate the matrix element of operator $\hat{M}_9$ in the spherical basis set
\begin{equation}\label{eqn:matrix-M9-sphere}
(\hat{M}_9)^{n,L,Q,J}_{ \lambda ^{\prime}, \lambda} = \langle \Psi ^{spherical}_{n,\lambda ^{\prime},L,Q,J,j_5,\ldots,j_1,m_j}| 
         \hat{M}_9 | \Psi ^{spherical}_{n,\lambda,L,Q,J,j_5,\ldots,j_1,m_j}\rangle ,
\end{equation} 
we use the connection between parabolic and spherical bases given in Ref. \cite{le2019para} as
\begin{equation}\label{eqn:sphere-para}
\Psi _{n, \lambda, L, Q, J, j_5, \ldots, j_1, m_j}^{spherical} = \sum _{n_p} W_{\lambda; n_p}^{n,L,Q,J} \Psi _{n, n_p, L, Q, J, j_5, \ldots, j_1, m_j}^{parabolic} .
\end{equation}
Here in \Eref{eqn:sphere-para}, matrix element $W_{\lambda; n_p}^{n,L,Q,J}$ of the interbasis transformation between spherical and parabolic bases was first given in Ref. \cite{le2019para} and is now written in a more compact form in \ref{app:A}. We can then easily transform the parabolic representation of $\hat{M}_9$ into the spherical representation by inserting \Eref{eqn:sphere-para} and \Eref{eqn:matrix-M9-para} into \Eref{eqn:matrix-M9-sphere} as
\begin{eqnarray}
(\hat{M}_9)^{n,L,Q,J}_{ \lambda ^{\prime}, \lambda} && = \sum _{n_p^{\prime},n_p} (\hat{M}_9)^{n,L,Q,J}_{ n_p^{\prime}, n_p} W_{\lambda ^{\prime}; n_p ^{\prime}}^{n,L,Q,J} W_{\lambda; n_p}^{n,L,Q,J} \nonumber\\
&& = - \sum _{n_p} \left( n + \frac{Q}{2} - J - 2 n_p \right) W_{\lambda ^{\prime}; n_p}^{n,L,Q,J} W_{\lambda; n_p}^{n,L,Q,J} \nonumber\\
&& = - \left( n + \frac{Q}{2} - J\right) \delta _{\lambda^{\prime} \lambda} + 2 \sum _{n_p} n_p W_{\lambda ^{\prime}; n_p}^{n,L,Q,J} W_{\lambda; n_p}^{n,L,Q,J} \label{eqn:matrix-M9-2}.
\end{eqnarray} 
The first term in the last line of \Eref{eqn:matrix-M9-2} comes from the orthogonality of matrix $W_{\lambda; n_p}^{n,L,Q,J}$, but the second term needs to be simplified. 

For this purpose, we develop a recurrence relation of matrix element $W_{\lambda; n_p}^{n,L,Q,J}$  and apply it to the second term of \Eref{eqn:matrix-M9-2}. Detailed calculations are given in \ref{app:A}. As a result, we obtain a compact form for this second term and consequently calculate the non-zero matrix elements of the Laplace-Runge-Lenz component $\hat{M}_9$ in the  spherical representation as
\begin{eqnarray}
(\hat{M}_9)^{n,L,Q,J}_{ \lambda ^{\prime}, \lambda} && =  \frac{(J-L)(L+J+6)(2n+Q+8)}{8 (\lambda + 3)(\lambda + 4)}\; \delta _{\lambda^{\prime} \lambda} \nonumber\\
&& \quad\quad\quad - B_{\lambda ^{\prime}} \; \delta _{\lambda^{\prime}-1, \lambda} - B_{\lambda} \;\delta _{\lambda^{\prime}+1, \lambda}  \label{eqn:matrix-M9},
\end{eqnarray}
with the coefficients
\begin{eqnarray}\label{eqn:B-rec}
&&  B_{\lambda} =  \sqrt{\left(n+\frac{Q}{2}-\lambda + 1\right) \left(n+\frac{Q}{2}+\lambda+7 \right)}   \nonumber\\
&& \times \sqrt{\frac{(\lambda - \frac{L+J}{2})(\lambda +6 + \frac{L+J}{2} ) (\lambda+3 - \frac{J-L}{2}) (\lambda+3 + \frac{J-L}{2})}{(\lambda+3)^2 (2\lambda+7)(2\lambda+5)}} .
\end{eqnarray}
This explicit expression \Eref{eqn:matrix-M9} of the matrix element is similar to that of the three- and five-dimensional MICZ-Kepler problems  \cite{Mardoyan2000,Mardoyan2003,coulson1967spheroidal}.
\subsection{Transformation between prolate spheroidal and spherical bases}\label{sec:4b}
Once the matrix elements of generalized angular momentum $\hat{\Lambda}^2$ and Laplace-Runge-Lenz component $\hat{M}_9$ are determined in the spherical representation, the eigenproblem by \Eref{eqn:spheroid} in prolate spheroidal coordinates can be solved using the spherical basis by the following interbasis transformation
\begin{equation}\label{eqn:T}
\Psi _{n, n_{k}, L, Q, J, j_5, \ldots, j_1, m_j}^{spheroidal} = \sum _{\lambda} T_{\lambda; n_{k}}^{n,L,Q,J} (a) \Psi _{n, \lambda, L, Q, J, j_5, \ldots, j_1, m_j}^{spherical}.
\end{equation}
Here, the coefficients must satisfy the normalization condition as
\begin{equation}\label{eqn:T-norm}
\sum _{\lambda} T_{\lambda; n_{k}^{\prime}}^{n,L,Q,J} (a) T_{\lambda; n_{k}}^{n,L,Q,J} (a) = \delta _{n_{k}^{\prime} n_{k}}
\end{equation} 
and have the natural limits
\begin{equation}\label{eqn:T-limit}
\left\{\begin{array}{l}
\lim _{a \to 0} T_{\lambda; n_{k}}^{n,L,Q,J} (a) = \delta _{n+Q/2-\lambda, n_{k}} \\
\lim _{a \to +\infty} T_{\lambda; n_{k}}^{n,L,Q,J} (a) = W_{\lambda; n - \frac{L+J-Q}{2} - n_{k}}^{n,L,Q,J}, 
\end{array}
\right. 
\end{equation}
as shown in Ref.\cite{le2019para}, since the prolate spheroidal coordinates become spherical or parabolic as $a \to 0$ or $a \to +\infty$, respectively. We note that the spherical solutions of \Eref{eqn:theta} give values of $\lambda$ from $(L+J)/2$ to $n+Q/2$. 

Substituting connection \Eref{eqn:T} into \Eref{eqn:spheroid} and using expressions \Eref{eqn:matrix-lambda} and \Eref{eqn:matrix-M9} for the matrix elements of operators $\hat{\Lambda}^2$ and $\hat{M}_9$ in the spherical basis, we get a set of linear equations governing the interbasis coefficients as a three-term recurrence relation
\begin{eqnarray}\label{eqn:K-solve} 
\left( A_{\lambda} - K  \right) T_{\lambda; n_{k}}^{n,L,Q,J} (a) 
 +  \tilde{B}_{\lambda}\; T_{\lambda - 1; n_{k}}^{n,L,Q,J} (a) + \tilde{B}_{\lambda + 1} \; T_{\lambda + 1; n_{k}}^{n,L,Q,J} (a) = 0,
\end{eqnarray}
where
\begin{eqnarray}\nonumber 
A_{\lambda} = - \frac{Z a (J-L)(L+J+6)}{4 (\lambda + 3) (\lambda +4)} - \lambda (\lambda + 7) ,\\\nonumber
 \tilde{B}_{\lambda} = \frac{2aZ}{2n+Q+8} {B}_{\lambda}.
\end{eqnarray}
For the index $\lambda$ running from $(L+J)/2$ to $n+Q/2$, we have a set of $N+1$ linear equations with $N= n+Q/2 - (L+J)/2$. \footnote{We use the definition of $N$ in analytical approach in Ref. \cite{le2019para} to keep notations consistent.} We can also rewrite them in a matrix form as
\begin{equation}\label{eqn:K-eqn-compact}
\left( \hat{\mathcal{K}} (a) - K \right) \hat{T} (a) = 0,
\end{equation}
where the solution is now presented in a column vector
$\hat{T} (a) = \left[ \ldots, T_{\lambda; n_{k}}^{n,L,Q,J}, \ldots \right]^{T}$; and $\hat{\mathcal{K}} (a)$ is a tridiagonal squared matrix of order $N+1$ as
\begin{equation}\label{eqn:spheroid-matrix}
\fl \hat{\mathcal{K}} (a) = \left[ \begin{array}{ccccc}
A_{\frac{L+J}{2}} & \tilde{B}_{\frac{L+J}{2}+1} & 0  & \ldots & 0\\
 \tilde{B}_{\frac{L+J}{2}+1} & A_{\frac{L+J}{2}+1}  &  \tilde{B}_{\frac{L+J}{2}+2}   & \ldots & 0\\
\vdots & \ddots & \ddots & \ddots & \vdots \\
0 & \ldots & 0 &  \tilde{B}_{n+\frac{Q}{2}} & A_{n+\frac{Q}{2}} 
\end{array}\right].
\end{equation}
We do not use the superscripts $n,L,Q,J$ for matrix $\hat{\mathcal{K}} (a)$ and column vector $\hat{T}(a)$ to avoid the cumbersomeness. 

For \Eref{eqn:K-eqn-compact}, the separation constant $K$ is the $n_k$-th eigenvalue of matrix $\hat{\mathcal{K}} (a)$. This value  depends on $a$ and can be calculated for fixed $n, L, Q, J$; thus, we denote the separation constant as $K_{n,n_k,L,Q,J}(a)$. Back to \Eref{eqn:K-solve}, fortunately, the similar three-term recurrence equations have been investigated in our previous works \cite{le2017,le2018}; thus, we can obtain a compact solution of these equations as follows. First, we introduce new coefficients
\begin{eqnarray}\nonumber
 t_{\lambda} (a) = T_{\lambda; n_{k}}^{n,L,Q,J} (a) \times (-1)^{\lambda - \frac{L+J}{2}} \prod _{j = \frac{L+J}{2}+1}^{\lambda} \tilde{B}_{j}.
\end{eqnarray}
Then multiply \Eref{eqn:K-solve} by the same factor to rewrite it as a new recurrence equation
\begin{eqnarray}\label{eqn:K-solve-2}
t_{\lambda+1} (a) = \left( A_{\lambda}  - K_{n,n_k,L,Q,J}(a)  \right) t_{\lambda} (a) - \tilde{B}_{\lambda}^2\;  t_{\lambda-1} (a)
\end{eqnarray}
with $\lambda$ running from $(L+J)/2$ to $n+Q/2$.
It is the well-known recurrence relation of a tridiagonal matrix determinant or so-called continuant \cite{Wilkinson65}. Hence $t_{\lambda} (a)$ can be expressed as a continuant
\begin{eqnarray}\label{eqn:T-solve}
t_{\lambda} (a) = C \det \left| \hat{\mathcal{K}}_{\lambda} (a) -  K \right|
\end{eqnarray}
with the normalization constant $C$ given in Ref.\cite{Wilkinson65} as
\begin{equation}
C = \left\{ \sum _{\lambda = \frac{L+J}{2}}^{n+\frac{Q}{2}} \prod _{j = \frac{L+J}{2}+1}^{\lambda} \tilde{B}_{j}^{-2}\;
 \det \left| \hat{\mathcal{K}}_{\lambda} (a) -  K \right| ^2 \right\} ^{-1/2}.
\end{equation}
Here, matrix $\hat{\mathcal{K}}_{\lambda} (a)$ is a tridiagonal matrix of order 
$N_{\lambda}=\lambda - \frac{L+J}{2}$ 
\begin{equation}
\fl \hat{\mathcal{K}}_{\lambda} (a) = \left[ \begin{array}{ccccc}
A_{\frac{L+J}{2}}  & \tilde{B}_{\frac{L+J}{2}+1} & 0   & \ldots & 0\\
 \tilde{B}_{\frac{L+J}{2}+1} & A_{\frac{L+J}{2}+1} &  \tilde{B}_{\frac{L+J}{2}+2}    & \ldots & 0\\
\vdots & \ddots & \ddots & \ddots & \vdots \\
0 & \ldots & 0 &  \tilde{B}_{\lambda-1} & A_{\lambda -1} 
\end{array}\right].
\end{equation}
Putting the separation constant $K _{n,n_k,L,Q,J} (a)$ into \Eref{eqn:T-solve}, we can calculate $t_{\lambda}(a)$ and the coefficient $T_{\lambda; n_{k}}^{n,L,Q,J} (a)$ consequently. This is a significant advantage of the algebraic approach where the separation constant $K_{n,n_{k},L,Q,J} (a)$ can principally be found just by calculating eigenvalues of the sparse matrix, and the explicit expression of the coefficient $T_{\lambda; n_{k}}^{n,L,Q,J} (a)$ can be determined by calculating the continuant.

When approaching this problem analytically in Ref. \cite{le2019para}, we could not calculate the coefficient $T_{\lambda; n_{k}}^{n,L,Q,J} (a)$ because of the complicated integral
\begin{eqnarray}\label{eqn:integral-T}
&& T_{\lambda; n_{k}}^{n,L,Q,J} (a) =  \int \dots \int r^8 dr \sin ^7 \theta d \theta d \Omega (\varphi , \phi )  \nonumber\\
&& \quad\quad\quad\times \Psi ^{spherical, *}_{n,\lambda,L,Q,J,{j_5},{j_4}{j_3},{j_2},{j_1},{m_j}} ( r , \theta , \varphi, \phi)  \nonumber\\
&& \quad\quad\quad\quad\times \Psi ^{spheroidal}_{n,n_{k},L,Q,J,{j_5},{j_4}{j_3},{j_2},{j_1},{m_j}} \left( \xi, \eta , \varphi,  \phi \right),
\end{eqnarray}
with the prolate spheroidal variables replaced by their expression in spherical coordinates. This integral contains three different types of orthogonal polynomial, namely confluently Heun, generalized Laguerre, generalized Jacobi, and on the best of our knowledge, has not been calculated elsewhere. Therefore, the compact expression of $T_{\lambda; n_{k}}^{n,L,Q,J} (a)$ conducted by the algebraic approach means that the integral \Eref{eqn:integral-T} is calculated. 


\subsection{Spherical and parabolic bases as limits of prolate spheroidal basis}\label{sec:4c}
Initially, the prolate spheroidal coordinates have two limits as spherical or parabolic regarding the distance $a$. Correspondingly, when taking the limits of $a \to 0$ or $a \to +\infty$, the prolate spheroidal basis of the 9D MICZ-KP must become the spherical basis or parabolic. This interesting property of the prolate spheroidal basis has been verified analytically by examining the prolate spheroidal wavefunction in Ref. \cite{le2019para}. Now we also give a proof by the algebraic perspective.  

First, we examine the limits of separation constant $K_{n,n_{k},L,Q,J} (a)$, which is the eigenvalue of the characterized equation (\ref{eqn:spheroid}) in prolate spheroidal coordinates. Upon the limits of the operator $\hat \Lambda_M^2=\hat \Lambda^2+a\sqrt{-2\hat H}\,\hat M_9$, \Eref{eqn:spheroid} becomes \Eref{eqn:sphere-define} when $a \to 0$ and \Eref{eqn:para} when $a \to +\infty$. In turn, these limits lead to the limit relations   
\begin{equation}\label{eqn:K-limit}
\fl \left\{ \begin{array}{ll}
\lim \limits_{a \to 0} K_{n,n_{k},L,Q,J} (a) &= - \left(n + \frac{Q}{2} - n_{k} \right) \left( n + \frac{Q}{2} - n_{k} + 7 \right) , \\
\lim \limits_{a \to +\infty} \frac{K_{n,n_{k},L,Q,J} (a)}{a} &= - \frac{2 Z \left( n + Q/2 - L - 2 n_{k} \right)}{2 n + Q + 8} ,
\end{array}
\right.
\end{equation}
corresponding to the separation constants of the spherical and parabolic bases, respectively.

Now, we consider the limit of $T_{\lambda; n_{k}}^{n,L,Q,J}(a)$ when $a \to 0$. In this case, all elements of the determinant become $A_{\lambda} (0) = - \lambda (\lambda + 7) $ and $\tilde{B}_{\lambda} (a) = 0$; thus, the continuant 
\begin{eqnarray}
&& \lim \limits_{a \to 0} \det \left| \hat{\mathcal{K}}_{\lambda} -  K_{n, n_{k}, L, Q, J} (a) \right|  \nonumber \\
&& = \prod _{j = \frac{L+J}{2}}^{\lambda -1} \left(n + Q/2 - n_{k} - j \right) \left( n + Q/2 - n_{k} + j + 7 \right)
\end{eqnarray}
vanishes for all $\lambda > n + \frac{Q}{2} - n_{k}$, and consequently, $T_{\lambda; n_{k}}^{n,L,Q,J}(a \to 0) = 0$ for this case. Conversely, when $\lambda \leq n + \frac{Q}{2} - n_{k}$, we pay attention to the normalization factor $C$, which has the dominant contribution at $\lambda = n + \frac{Q}{2} - n_{k}$ so that 
$T_{\lambda; n_{k}}^{n,L,Q,J}(a \to 0) \sim a^{n+Q/2 - n_{k} - \lambda}$. As a result, we have
\begin{eqnarray}\label{eqn:11a}
\lim \limits _{a \to 0} T_{\lambda; n_{k}}^{n,L,Q,J} (a) 
&&  =  \left\{ \begin{array}{ll}
0 & \mathrm{if \; } \lambda \neq n+Q/2 - n_{k} \\
1 & \mathrm{if \; } \lambda = n+Q/2 - n_{k} 
\end{array} \right. \nonumber\\
 && = \delta _{n+Q/2 -\lambda , n_{k}}.
\end{eqnarray} 

On the other hand, when $a \to +\infty$, the three-term recurrence relation \Eref{eqn:K-solve} approximately becomes 
\begin{eqnarray}
\fl  \left\{ \frac{(J-L)(L+J+6)(2n+Q+8)}{8 (\lambda + 3) (\lambda +4)} +  2n_{k} + n + \frac{Q}{2} - L  \right\} T_{\lambda; n_{k}}^{n,L,Q,J} (+\infty) \nonumber\\
\fl \quad \quad \quad \quad \quad \quad  \quad \quad \quad \quad \quad \quad \quad \quad \quad  -  B_{\lambda} T_{\lambda - 1; n_{k}}^{n,L,Q,J} (+\infty) - B_{\lambda + 1} T_{\lambda + 1; n_{k}}^{n,L,Q,J} (+\infty) \approx 0,
\end{eqnarray}
identical to the three-term recurrence relation \Eref{eqn:W-rec} of $W_{\lambda; n_p}^{n,L,Q,J}$ with $n_p = n - \frac{L+J-Q}{2} - n_{k} $. Hence, we can conclude that
\begin{equation}\label{eqn:11b}
\lim\limits _{a \to +\infty} T_{\lambda; n_{k}}^{n,L,Q,J} (a) = W_{\lambda; n - \frac{L+J-Q}{2} - n_{k}}^{n,L,Q,J}.
\end{equation}
\Eref{eqn:11a} and \Eref{eqn:11b} are consistent with the results given in \Eref{eqn:T-limit}.

\section{Conclusion}\label{sec:conc}
In this study, we have explicitly constructed three versions for the last member of the nonuplet of algebraically independent integrals of motion for the nine-dimensional MICZ-Kepler problem and related them to the variable separation of this problem. Particularly, we have shown that the last constant of motion can be either the total generalized angular momentum $\hat \Lambda^2$ or the last component of the Laplace-Runge-Lenz vector $\hat M_9$, or their combination $\hat \Lambda^2_M=\hat \Lambda^2+ a\sqrt{-2\hat H} \hat M_9$ corresponding to the spherical, parabolic, or prolate spheroidal bases. Especially, we have variable-separated the Schr{\"o}dinger equation in spherical, parabolic, and prolate spheroidal coordinates and proved that the separation constants are the chosen integrals of motion.  

Another significant result is the relation between the prolate spheroidal and spherical bases. We have established the interbasis transformation coefficients between the bases by using the algebraic structure behind these bases. 
Furthermore, we have shown an algebraic approach to calculate the nine-dimensional complicated integrals of confluent Heun, generalized Laguerre, and Jacobi polynomials using the results above.

\section*{Acknowledgement}
We thank Professor Pinaki Roy (Atomic Molecular and Optical Physics Research Group, Advanced Institute of Materials Science, Ton Duc Thang University) for his helpful comments when finalizing this work. Dai-Nam Le was funded by Vingroup Joint Stock Company and supported by the Domestic Master/ PhD Scholarship Programme of Vingroup Innovation Foundation (VINIF), Vingroup Big Data Institute (VINBIGDATA), code VINIF.2020.TS.03.

\section*{Author contribution statement}

All authors contributed equally to the paper. All the authors have read and approved the final manuscript.

\section*{References}
\bibliographystyle{iopart-num} \bibliography{MICZ}

\providecommand{\newblock}{}
\begin{thebibliography}{10}
\expandafter\ifx\csname url\endcsname\relax
  \def\url#1{{\tt #1}}\fi
\expandafter\ifx\csname urlprefix\endcsname\relax\def\urlprefix{URL }\fi
\providecommand{\eprint}[2][]{\url{#2}}

\bibitem{zwanziger1968exactly}
Zwanziger D 1968 {\em Physical Review\/} {\bf 176} 1480

\bibitem{mcintosh1970degeneracy}
McIntosh H~V and Cisneros A 1970 {\em Journal of Mathematical Physics\/} {\bf
  11} 896--916

\bibitem{Dirac1931}
Dirac P~A~M 1931 {\em Proceedings of the Royal Society A: Mathematical,
  Physical and Engineering Sciences\/} {\bf 133} 60--72

\bibitem{Nieto1979}
Nieto M~M 1979 {\em American Journal of Physics\/} {\bf 47} 1067--1072

\bibitem{MLADENOV1985}
Mladenov I and Tsanov V 1985 {\em Journal of Geometry and Physics\/} {\bf 2} 17
  -- 24

\bibitem{AQUILANTI1997}
Aquilanti V, Cavalli S and Coletti C 1997 {\em Chemical Physics\/} {\bf 214} 1
  -- 13

\bibitem{al1998hydrogen}
Al-Jaber S~M 1998 {\em International Journal of Theoretical Physics\/} {\bf 37}
  1289--1298

\bibitem{Iwai1990}
Iwai T 1990 {\em Journal of Geometry and Physics\/} {\bf 7} 507--535 ISSN
  03930440

\bibitem{Iwai1996}
Iwai T and Sunako T 1996 {\em Journal of Geometry and Physics\/} {\bf 20}
  250--272 ISSN 03930440

\bibitem{Mardoyan1997}
Mardoyan L~G, Sissakian A~N and Ter-Antonyan V~M 1997  ISSN 10637788
  (\textit{Preprint} \eprint{9712235})

\bibitem{le2009hidden}
Le V~H, Nguyen T~S and Phan N~H 2009 {\em Journal of Physics A: Mathematical
  and Theoretical\/} {\bf 42} 175204

\bibitem{le2011non}
Le V~H and Nguyen T~S 2011 {\em Journal of Mathematical Physics\/} {\bf 52}
  032105

\bibitem{Meng2007}
Meng G 2007 {\em Journal of Mathematical Physics\/} {\bf 48} 032105 ISSN
  0022-2488

\bibitem{Meng2008}
Meng G 2008 {\em Physics of Atomic Nuclei\/} {\bf 71} 946--950 ISSN 1063-7788

\bibitem{Meng2010}
Meng G 2010 {\em Journal of the London Mathematical Society\/} {\bf 81}
  663--678 ISSN 00246107

\bibitem{Meng2011}
Meng G and Zhang R 2011 {\em Journal of Mathematical Physics\/} {\bf 52} 042106

\bibitem{Barut1979}
Barut A~O, Schneider C~K~E and Wilson R 1979 {\em Journal of Mathematical
  Physics\/} {\bf 20} 2244--2256

\bibitem{KS1965}
Kustaanheimo P and Stiefel E 1965 {\em Journal f\"{u}r die reine und angewandte
  Mathematik (Crelles Journal)\/} {\bf 218} 204--219

\bibitem{kibler1986hydrogen}
Kibler M, Ronveaux A and Négadi T 1986 {\em Journal of Mathematical Physics\/}
  {\bf 27} 1541--1548

\bibitem{davtyan1987generalized}
Davtyan L, Mardoyan L, Pogosyan G, Sissakian A and Ter-Antonyan V 1987 {\em
  Journal of Physics A: Mathematical and General\/} {\bf 20} 6121

\bibitem{Le1991}
Le V~H, Viloria T~J and Le A~T 1991 {\em Journal of Physics A: Mathematical and
  General\/} {\bf 24} 3021--3030 ISSN 03054470

\bibitem{Le1993}
Le V~H and Komarov L~I 1993 {\em Physics Letters A\/} {\bf 177} 121--124 ISSN
  03759601

\bibitem{Yang1978}
Yang C~N 1978 {\em Journal of Mathematical Physics\/} {\bf 19} 320--328 ISSN
  0022-2488

\bibitem{grossman1984solutions}
Grossman B, Kephart T~W and Stasheff J~D 1984 {\em Communications in
  Mathematical Physics\/} {\bf 96} 431--437

\bibitem{pedder2008berry}
Pedder C, Sonner J and Tong D 2008 {\em Journal of High Energy Physics\/} {\bf
  2008} 065

\bibitem{baez2002octonions}
Baez J 2002 {\em Bulletin of the American Mathematical Society\/} {\bf 39}
  145--205

\bibitem{hopf1931abbildungen}
Hopf H 1931 {\em Mathematische Annalen\/} {\bf 104} 637--665

\bibitem{hopf1935abbildungen}
Hopf H 1935 {\em Fundamenta Mathematicae\/} {\bf 25} 427--440

\bibitem{Laughlin1983}
Laughlin R~B 1983 {\em Phys. Rev. Lett.\/} {\bf 50}(18) 1395--1398

\bibitem{zhang2001four}
Zhang S~C and Hu J 2001 {\em Science\/} {\bf 294} 823--828

\bibitem{bernevig2003eight}
Bernevig B~A, Hu J, Toumbas N and Zhang S~C 2003 {\em Physical Review
  Letters\/} {\bf 91} 236803

\bibitem{FUREY2018}
Furey C 2018 {\em Physics Letters B\/} {\bf 785} 84 -- 89 ISSN 0370-2693

\bibitem{Boyle2020}
Boyle L and Farnsworth S 2020 {\em New Journal of Physics\/} {\bf 22} 073023

\bibitem{Mardoyan1999}
Mardoyan L~G, Sissakian A~N and Ter-Antonyan V~M 1999 {\em Modern Physics
  Letters A\/} {\bf 14} 1303--1307 ISSN 0217-7323

\bibitem{Mardoyan2000}
Mardoyan L~G, Sisakyan A~N and Ter-Antonyan V~M 2000 {\em Theoretical and
  Mathematical Physics\/} {\bf 123} 451--462 ISSN 0040-5779

\bibitem{Mardoyan2003}
Mardoyan L 2003 {\em Journal of Mathematical Physics\/} {\bf 44} 4981--4987
  ISSN 0022-2488

\bibitem{Pletyukhov1999}
Pletyukhov M~V and Tolkachev E~A 1999 {\em Journal of Physics A: Mathematical
  and General\/} {\bf 32} L249--L253 ISSN 0305-4470

\bibitem{Marquette2010}
Marquette I 2010 {\em Journal of Mathematical Physics\/} {\bf 51} 102105 ISSN
  0022-2488

\bibitem{Marquette2012}
Marquette I 2012 {\em Journal of Mathematical Physics\/} {\bf 53} 022103 ISSN
  0022-2488

\bibitem{Hoque2017}
Hoque M~F, Marquette I and Zhang Y~Z 2017 {\em Annals of Physics\/} {\bf 380}
  121--134 ISSN 00034916

\bibitem{le2011so}
Le V~H, Phan T~T and Truong C~T 2011 {\em Journal of Mathematical Physics\/}
  {\bf 52} 072101

\bibitem{phan2012generalized}
Phan N~H and Le V~H 2012 {\em Journal of Mathematical Physics\/} {\bf 53}
  082103

\bibitem{nguyen2015exact}
Nguyen T~S, Le D~N, Thoi T~Q~N and Le V~H 2015 {\em Journal of Mathematical
  Physics\/} {\bf 56} 052103

\bibitem{le2019para}
Le D~N, Phan N~H, Thoi T~Q~N and Le V~H 2019 {\em Journal of Mathematical
  Physics\/} {\bf 60} 062102

\bibitem{phan2018super}
Phan N~H, Le D~N, Thoi T~Q~N and Le V~H 2018 {\em Journal of Mathematical
  Physics\/} {\bf 59} 032102

\bibitem{Kereselidze2016}
Kereselidze T, Chkadua G and Ogilvie J~F 2016 {\em GESJ: Physics\/} {\bf 2}
  44--56

\bibitem{gradshteyn2014table}
Gradshteyn I~S and Ryzhik I~M 2014 {\em Table of integrals, series, and
  products\/} (Academic press)

\bibitem{Mardoyan2005}
Mardoyan L~G 2005 {\em Physics of Atomic Nuclei\/} {\bf 68} 1746--1755 ISSN
  1063-7788

\bibitem{coulson1967spheroidal}
Coulson C~A and Joseph A 1967 {\em Proceedings of the Physical Society\/} {\bf
  90} 887

\bibitem{le2017}
Le D~N, Hoang N~T~D and Le V~H 2017 {\em Journal of Mathematical Physics\/}
  {\bf 58} 042102

\bibitem{le2018}
Le D~N, Hoang N~T~D and Le V~H 2018 {\em Journal of Mathematical Physics\/}
  {\bf 59} 032101

\bibitem{Wilkinson65}
Wilkinson J~H 1965 {\em Algebraic Eigenvalue Problem\/} (New York: Oxford
  University Presss)

\bibitem{Tarter1970}
Tarter C~B 1970 {\em Journal of Mathematical Physics\/} {\bf 11} 3192--3195
  ISSN 0022-2488

\bibitem{Rainville1945}
Rainville E~D 1945 {\em Bulletin of the American Mathematical Society\/} {\bf
  51} 714--724 ISSN 0002-9904

\bibitem{Varshalovich1988}
Varshalovich D~A, Moskalev A~N and Khersonskii V~K 1988 {\em Quantum Theory of
  Angular Momentum\/} (WORLD SCIENTIFIC)

\end{thebibliography}

\appendix

\section{Interbasis transformation between parabolic and spherical bases}\label{app:A}
For further calculations, we recall the explicit spherical and parabolic bases given in Refs. \cite{nguyen2015exact,le2019para} as
\begin{eqnarray}
\fl \Psi _{n, \lambda, L, Q, J, j_5,\ldots, j_1, m_j}^{spherical} (r, \theta, \varphi, \phi) = \psi _{n, \lambda, L,Q,J}^{spherical} (r, \theta) \mathcal{D}_{L, Q, J, j_5,\ldots, j_1, m_j} (\varphi, \phi), \\
\fl \Psi _{n, n_p, L, Q, J, j_5,\ldots, j_1, m_j}^{parabolic} (u, v, \varphi, \phi) = \psi _{n, n_p, L, Q, J}^{parabolic} (u, v) \mathcal{D}_{L, Q, J, j_5,\ldots, j_1, m_j} (\varphi, \phi),
\end{eqnarray}
in which
\begin{eqnarray}\label{eqn:11A}
\fl \psi _{n, \lambda, L,Q,J}^{spherical} (r, \theta) = &&  C_{n,\lambda, L,Q,J}^{s} \alpha ^{9/2}  \nonumber\\
 &&  \times \left( \alpha r \right)^{\lambda} \exp{ \left( - \alpha r / 2 \right)} \;\mathcal{L}_{n+Q/2 - \lambda}^{2 \lambda + 7} \left( \alpha r \right) \nonumber \\
 &&  \times 2^{-(L+J+7)/2} \left(1 - \cos \theta \right) ^{L/2} \left(1 + \cos \theta \right) ^{J/2} \mathcal{P}_{\lambda - (L+J)/2}^{(L+3,J+3)} (\cos \theta) ,\\
\fl \psi _{n, n_p, L, Q, J}^{parabolic} (u, v) = && C_{n,n_p, L,Q,J}^{p} 2^{-7/2} \alpha ^{9/2} \nonumber\\
 &&  \times \left( \alpha u / 2 \right)^{J/2} \exp{ \left( - \alpha u / 4 \right)}\; \mathcal{L}_{n_p}^{J+3} \left( \alpha u / 2 \right)  \nonumber \\
 &&  \quad \quad \quad \times \left( \alpha v / 2 \right)^{L/2} \exp{ \left( - \alpha v / 4 \right)}\; \mathcal{L}_{n+Q/2 - (L+J)/2 - n_p}^{L+3} \left( \alpha v / 2 \right).
\end{eqnarray}
Here, $\mathcal{L}$ and $\mathcal{P}$ are the generalized Laguerre and Jacobi polynomials whose definitions are given in Ref. \cite{gradshteyn2014table}; $\alpha = 4Z (2n+Q+8)^{-1}$; and $C_{n,\lambda, L,Q,J}^{s}$, $C_{n,n_p, L,Q,J}^{p}$ are normalization constants given as follows
\begin{eqnarray}
\fl C_{n,\lambda, L,Q,J}^{s} && = \sqrt{ \frac{(n+\frac{Q}{2} - \lambda)! \,(2\lambda + 7) (\lambda - \frac{L+J}{2})! \,(\lambda +6 +\frac{L+J}{2})!}{(2n + Q + 8) (n+\frac{Q}{2}+\lambda + 7)!(\lambda +3 + \frac{J-L}{2} )! (\lambda+3 - \frac{J-L}{2})! }}\,, \\
\fl C_{n,n_p, L,Q,J}^{p} && = \sqrt{\frac{n_p! (n+\frac{Q}{2} - \frac{L+J}{2}  - n_p)! }{ (2 n + Q + 8) (n_p+J+3)! (n+\frac{Q}{2} + \frac{J-L}{2}  - n_p + 3)! }}\,.
\end{eqnarray}
Then the interbasis transformation between the spherical and parabolic bases of the 9D MICZ-KP is defined as 
\begin{equation}
\Psi _{n, \lambda, L, Q, J, j_5, \ldots, j_1, m_j}^{spherical} = \sum _{n_p} W_{\lambda; n_p}^{n,L,Q,J} \Psi _{n, n_p, L, Q, J, j_5, \ldots, j_1, m_j}^{parabolic}
\end{equation}
with coefficients determined by the following integrals
\begin{equation} \label{eqn:W-integral}
W_{\lambda; n_p}^{n,L,Q,J} = \int \psi _{n, n_p, L, Q, J}^{parabolic, *} (u, v) \psi _{n, \lambda, L,Q,J}^{spherical} (r, \theta) r^{8}\sin^7 \theta dr d \theta ,
\end{equation}
derived in Ref. \cite{le2019para} for the first time. In this Appendix, we will represent this interbasis in a more compact form as a recurrence relation, which is necessary for application in Section \ref{sec:4}. 

Plugging wavefunctions \Eref{eqn:11A} into \Eref{eqn:W-integral} and doing some expansions, we obtain the following formula
\begin{eqnarray}\label{eqn:W-sum}
&& W_{\lambda; n_p}^{n,L,Q,J} = C_{n,\lambda, L,Q,J}^{s} C_{n,n_p, L,Q,J}^{p}   \nonumber\\
&& \quad \quad \times\sum _{s = 0}^{n_p} \sum _{t = 0}^{n+\frac{Q}{2} - (L+J)/2 - n_p} \biggl\{ \frac{(-1)^{s+t}}{s! t!} (n_p-s+1)_{(s+J+3)} \biggr. \nonumber\\
&& \quad \quad\quad \quad \times \biggl. \left(n+\frac{Q}{2} - \frac{L+J}{2}  - n_p - t + 1 \right)_{(t+L+3)} I_{s,t} K_{s, t} \biggr\}
\end{eqnarray}
with the Pochhammer rising factorial $(a)_{(b)} = a (a+1) \ldots (a+b-1) $ and two integrals calculated in Ref. \cite{gradshteyn2014table} as
\begin{eqnarray}
\fl I_{s,t} && = \int _0^{+\infty} e^{-x} x^{\lambda + \frac{L+J}{2} + 8 + s + t} \mathcal{L}_{n+{Q}/{2}-\lambda}^{2\lambda + 7} (x) dx \nonumber \\
\fl && = \frac{(\lambda + \frac{L+J}{2} + 8 + s + t)!}{(n+{Q}/{2} - \lambda)!} \left( \lambda - \frac{L+J}{2} - s - t -1 \right) _{(n+{Q}/{2} - \lambda)} , \label{eqn:I-integral} \\
\fl K_{s,t} && = \frac{2^{-(L+J+7+s+t)}}{(s+J+3)! (t+L+3)!} \int _{-1}^{+1} (1-x)^{t+L+3} (1+x)^{s+J+3} \mathcal{P}_{\lambda - \frac{L+J}{2}}^{(L+3,J+3)} (x) dx \nonumber \\
\fl && = \frac{(\lambda - \frac{J-L}{2} + 3)!}{(L+3)! (\lambda + \frac{J-L}{2})! (L+J+s+t+7)!}  \nonumber\\
\fl && \quad \quad \quad \times {}_{3} F_{2} \left[ \begin{array}{c} -(\lambda - \frac{L+J}{2}), \lambda + \frac{L+J}{2} + 7, t+L+4\\
 L+4, L+J+s+t+8 \end{array} ; 1\right] \nonumber\\
\fl && = \frac{\left(\lambda - \frac{J-L}{2}+ 3\right)!}{(L+3)! \left(\lambda - \frac{L+J}{2}\right)! (L+J+s+t+7)!} \frac{\left[ -(s+t) \right]_{\left(\lambda - \frac{L+J}{2}\right)}}{\left[ -(\lambda + \frac{L+J}{2} + 7+s+t) \right]_{\left( \lambda - \frac{L+J}{2} \right) }}  \nonumber\\
\fl && \quad \quad \quad \times {}_{3} F_{2} \left[ \begin{array}{c} -\left(\lambda - \frac{L+J}{2}\right), - t, \lambda + \frac{L+J}{2} + 7 \\
 L+4, -(s+t) \end{array} ; 1\right]. \label{eqn:K-integral}
\end{eqnarray}   
 
As can be seen from \Eref{eqn:K-integral}, integral $K_{s,t}$ non vanishes if and only if $\lambda - (L+J)/2 - (s+t) -1 < 0$. Therefore \Eref{eqn:I-integral} should be rewritten as
\begin{eqnarray}\label{eqn:I-integral-2} 
I_{s,t} = && \frac{\left(\lambda + \frac{L+J}{2} + 8 + s + t \right)!}{\left(n+{Q}/{2} - \lambda \right)!} \nonumber\\
&& \times \left[ - \left( (s+t) - \left(\lambda - \frac{L+J}{2} \right) +1 \right) \right] _{\left(n+ {Q}/{2} - \lambda \right)} ,
\end{eqnarray}
which is only nonzero when $\left( s+t - (\lambda - L/2 - J/2 ) +1 \right) < n + Q/2 - \lambda - 1 \Rightarrow s+t \geq n + Q/2 -L/2 - J/2 - 1$. Comparing with the boundary values of  indices $(s,t)$, the nonzero terms in \Eref{eqn:W-sum} correspond to   
\begin{equation}
\left\{ \begin{array}{cll}
(1) & s = n_p, & t = n + \frac{Q}{2} - \frac{L+J}{2} - n_p, \\
(2) & s = n_p, & t = n + \frac{Q}{2} - \frac{L+J}{2} - n_p - 1, \\
(3) & s = n_p-1 , & t = n + \frac{Q}{2} - \frac{L+J}{2} - n_p .
\end{array} \right.
\end{equation}
Replacing these expressions into \Eref{eqn:W-sum}, then reducing the last two terms $(2)$ and $(3)$ by the contiguous relation for ${}_{3} F_{2} (z = 1)$ given in \cite{Tarter1970,Rainville1945} as
\begin{eqnarray}
\fl  n_p \; {}_{3} F_{2} \left[ \begin{array}{c} -(\lambda - \frac{L+J}{2}), - (n+\frac{Q}{2} - \frac{L+J}{2})+n_p, \lambda + \frac{L+J}{2} + 7 \\
 L+4, -\left(n+\frac{Q}{2} - \frac{L+J}{2}\right)+1 \end{array} ; 1\right] \nonumber\\ 
\fl \quad \quad  = 
- \left( n+\frac{Q}{2} - \frac{L+J}{2} - n_p \right) \nonumber\\
\fl  \quad \quad \quad \quad \quad \times {}_{3} F_{2} \left[ \begin{array}{c} -\left( \lambda - \frac{L+J}{2} \right), - \left(n+\frac{Q}{2} - \frac{L+J}{2} \right)+n_p + 1, \lambda + \frac{L+J}{2} + 7 \\
 L+4, -\left( n+\frac{Q}{2} - \frac{L+J}{2}\right)+1 \end{array} ; 1\right]
 \nonumber\\
\fl \quad \quad \quad + \left( n+\frac{Q}{2} - \frac{L+J}{2} \right)  \nonumber\\
\fl \quad \quad \quad \quad  \quad \times {}_{3} F_{2} \left[ \begin{array}{c} -\left(\lambda - \frac{L+J}{2}\right), - \left(n+\frac{Q}{2} - \frac{L+J}{2}\right)+n_p, \lambda + \frac{L+J}{2} + 7 \\
 L+4, -\left(n+\frac{Q}{2} - \frac{L+J}{2} \right) \end{array} ; 1\right],
\end{eqnarray}
followed by comparing them with the ${}_{3} F_{2} (z = 1)$-form of $SU(2)$ Clebscb-Gordan coefficients (see in Refs.  \cite{Varshalovich1988,Mardoyan2000}), we finally obtain the compact form of the coefficients
\begin{eqnarray}\label{eqn:W-compact}
\fl  W_{\lambda; n_p}^{n,L,Q,J} && = (-1)^{\lambda -(L+J)/2} \frac{(n+\frac{Q}{2}- \frac{L+J}{2})!}{(L+3)!} \sqrt{ \frac{(\lambda + \frac{L+J}{2} + 6)!}{ (\lambda - \frac{L+J}{2})!}} \nonumber \\ 
\fl  &&  \quad \quad  \times \sqrt{\frac{(2\lambda+7) (\lambda - \frac{J-L}{2} +3)!}{(n+\frac{Q}{2}+\lambda +7)! (n+\frac{Q}{2} - \lambda)! (\lambda +\frac{J-L}{2} +3)! }}  \nonumber\\
\fl  &&  \quad \quad \quad \times \sqrt{\frac{(n_p+J+3)!}{(n+\frac{Q}{2}- \frac{L+J}{2} - n_p)!} \frac{(n+\frac{Q}{2}- \frac{L+J}{2} - n_p + L+3))!}{n_p!} } \nonumber\\
\fl  &&  \quad \quad \quad \quad \times  {}_{3} F_{2} \left[ \begin{array}{c} -(\lambda - \frac{L+J}{2}), - (n+\frac{Q}{2} -\frac{L+J}{2}) + n_p, \lambda + \frac{L+J}{2} + 7 \\
 L+4, -(n+\frac{Q}{2} - \frac{L+J}{2}) \end{array} ; 1\right] \nonumber\\
\fl  &&  = (-1)^{n + {Q}/{2}-\lambda - n_p} \nonumber\\
\fl  &&  \times \mathcal{C}_{\frac{n+3}{2}+\frac{Q+J-L}{4},n_p-\frac{n}{2}+\frac{L+J-Q}{4}+\frac{J+3}{2}; \frac{n+3}{2}+\frac{Q-J+L}{4} ,\frac{n}{2}-\frac{L+J-Q}{4} - n_p +\frac{L+3}{2}}^{\lambda+3, \frac{L+J}{2}+3} .
\end{eqnarray}
The above expression has the same form as the Tarter coefficient of the three-dimensional hydrogen atom \cite{Tarter1970} and also consistent with that of three- and five-dimensional MICZ-Kepler problem  \cite{Mardoyan2000,Mardoyan2005}.

Now utilizing the well-known recurrence relation of $SU(2)$ Clebscb-Gordan coefficients \cite{Varshalovich1988}
\begin{eqnarray}
\fl \left\{ \frac{\alpha - \beta}{2} - \frac{\gamma (a-b)(a+b+1)}{2 c (c+1)} \right\} \mathcal{C}_{a,\alpha; b,\beta}^{c,\gamma} = \nonumber\\
\fl - \sqrt{\frac{(c-\gamma+1)(c+\gamma+1)(b-a+c+1)(a-b+c+1)(a+b-c)(a+b+c+2)}{4 (c+1)^2 (2c+3) (2c+1)}} \mathcal{C}_{a,\alpha; b,\beta}^{c+1,\gamma} \nonumber\\
\fl - \sqrt{\frac{(c-\gamma)(c+\gamma)(b-a+c)(a-b+c)(a+b-c+1)(a+b+c+1)}{4 c^2 (2c+1) (2c-1)}} \mathcal{C}_{a,\alpha; b,\beta}^{c-1,\gamma} ,
\end{eqnarray}
we finally show that the coefficients must obey the following recurrence relation
\begin{eqnarray}\label{eqn:W-rec}
\fl  \left[n_p- \frac{1}{2} \left( n + {Q}/{2} - J \right) - \frac{(J-L)(L+J+6)(2n+Q+8)}{16 (\lambda + 3)(\lambda + 4)} \right] W_{\lambda; n_p}^{n,L,Q,J}  \nonumber\\
\fl  \quad \quad \quad \quad \quad \quad \quad \quad \quad \quad \quad \quad = - \frac{1}{2} \left( B_{\lambda} W_{\lambda - 1; n_p}^{n,L,Q,J} + B_{\lambda+1} W_{\lambda+1; n_p}^{n,L,Q,J} \right),
\end{eqnarray}
with 
\begin{eqnarray}
\fl B_{\lambda} = \sqrt{(n+{Q}/{2}-\lambda + 1) (n+{Q}/{2}+\lambda+7)} \nonumber\\
\fl \quad\quad\quad\quad \times \sqrt{\frac{(\lambda - \frac{L+J}{2})(\lambda + \frac{L+J}{2} + 6) (\lambda - \frac{J-L}{2}+ 3) (\lambda + \frac{J-L}{2}+ 3)}{(\lambda+3)^2 (2\lambda+7)(2\lambda+5)}} .
\end{eqnarray}


Furthermore, in \Eref{eqn:matrix-M9-2}, the last term $2 \sum _{n_p} n_p W_{\lambda ^{\prime}; n_p}^{n,L,Q,J} W_{\lambda; n_p}^{n,L,Q,J}$ needs to be simplified so that the matrix element of Laplace-Runge-Lentz component $\hat{M}_9$ in the spherical representation is determined as \Eref{eqn:matrix-M9}. This simplification is as follows. First, we rewrite the recurrence relation \Eref{eqn:W-rec}:
\begin{eqnarray}
\fl  2n_p W_{\lambda ^{\prime}; n_p}^{n,L,Q,J} = \left[ n + {Q}/{2} - J  + \frac{(J-L)(L+J+6)(2n+Q+8)}{8 (\lambda^{\prime} + 3)(\lambda^{\prime} + 4)} \right] W_{\lambda^{\prime}; n_p}^{n,L,Q,J}  \nonumber\\
\fl  \quad \quad \quad \quad \quad \quad \quad \quad - B_{\lambda^{\prime}} W_{\lambda^{\prime} - 1; n_p}^{n,L,Q,J} - B_{\lambda^{\prime}+1} W_{\lambda^{\prime}+1; n_p}^{n,L,Q,J} .
\end{eqnarray}
Then replace the term $2 n_p W_{\lambda ^{\prime}; n_p}^{n,L,Q,J}$ in the sum $2 \sum _{n_p} n_p W_{\lambda ^{\prime}; n_p}^{n,L,Q,J} W_{\lambda; n_p}^{n,L,Q,J}$ by the right-hand-side of the above equation to obtain 
\begin{eqnarray}
\fl  \sum _{n_p} n_p W_{\lambda ^{\prime}; n_p}^{n,L,Q,J} W_{\lambda; n_p}^{n,L,Q,J} = \nonumber\\
\fl  \quad \quad \sum _{n_p} \left\{\left[ n + {Q}/{2} - J  + \frac{(J-L)(L+J+6)(2n+Q+8)}{8 (\lambda^{\prime} + 3)(\lambda^{\prime} + 4)} \right] W_{\lambda^{\prime}; n_p}^{n,L,Q,J} W_{\lambda; n_p}^{n,L,Q,J} \right. \nonumber\\
\fl  \quad \quad \quad \quad \left. -  B_{\lambda^{\prime}} W_{\lambda^{\prime} - 1; n_p}^{n,L,Q,J} W_{\lambda; n_p}^{n,L,Q,J} - B_{\lambda^{\prime}+1} W_{\lambda^{\prime}+1; n_p}^{n,L,Q,J} W_{\lambda; n_p}^{n,L,Q,J} \right\} .
\end{eqnarray}
Now taking into account the orthogonality of the matrix $W_{\lambda; n_p}^{n,L,Q,J}$, we finally reduce the explicit expression of the sum as the form of a tridiagonal matrix
\begin{eqnarray}
 && \quad\quad 2 \sum _{n_p} n_p W_{\lambda ^{\prime}; n_p}^{n,L,Q,J} W_{\lambda; n_p}^{n,L,Q,J} \nonumber\\
 && \quad\quad\quad \quad = \left[ n + {Q}/{2} - J  + \frac{(J-L)(L+J+6)(2n+Q+8)}{8 (\lambda + 3)(\lambda + 4)} \right] \delta _{\lambda ^{\prime} \lambda } \nonumber\\
 && \quad\quad\quad \quad \quad \quad -  B_{\lambda+1} \delta _{\lambda ^{\prime}-1, \lambda } - B_{\lambda} \delta _{\lambda ^{\prime}+1, \lambda} ,
\end{eqnarray}
from which we determine the matrix element of Laplace-Runge-Lentz component $\hat{M}_9$ in the spherical representation given in \Eref{eqn:matrix-M9}.
\end{document}